\documentclass[preprint,nolinenumbers]{aastex631}

\usepackage{graphicx,latexsym} 
\usepackage{amssymb,amsmath}

\usepackage{hyperref}
\usepackage{multirow}

\submitjournal{ApJ}

\begin{document}

\title{Compression Acceleration of Protons and Heavier Ions at the Heliospheric Current Sheet}

\correspondingauthor{Giulia Murtas}
\email{giuliamurtas31994@gmail.com}

\author[0000-0002-7836-7078]{Giulia Murtas}
\affiliation{Los Alamos National Laboratory,
Los Alamos, NM 87545, USA}

\author[0000-0001-5278-8029]{Xiaocan Li}
\affiliation{Dartmouth College,
Hanover, NH 03755, USA}

\author[0000-0003-4315-3755]{Fan Guo}
\affiliation{Los Alamos National Laboratory, 
Los Alamos, NM 87545, USA}

\begin{abstract}

Recent observations by Parker Solar Probe (PSP) suggest that protons and heavier ions are accelerated to high energies by magnetic reconnection at the heliospheric current sheet (HCS). By solving the energetic particle transport equation in large-scale MHD simulations, we study the compression acceleration of protons and heavier ions in the reconnecting HCS. We find that the acceleration of multi-species ions results in nonthermal power-law distributions with spectral index consistent with the PSP observations. Our study shows that the high-energy cutoff of protons can reach $E_{max} \sim 0.1$ -- $1$ MeV depending on the particle diffusion coefficients. We also study how the high-energy cutoff of different ion species scales with the charge-to-mass ratio  $E_{max} \propto (Q/M)^\alpha$. When determining the diffusion coefficients from the quasilinear theory with a Kolmogorov magnetic power spectrum, we find that $\alpha \sim 0.4$, which is somewhat smaller than $\alpha \sim 0.7$ observed by PSP.

\end{abstract}

\keywords{Magnetic reconnection --- Plasma physics --- Particle acceleration --- Heliospheric physics}

\section{Introduction} \label{sec:intro}

The heliospheric current sheet (HCS) is a structure originating from the outward extension of the solar magnetic dipole \citep{Smith2001}. 
Recent Parker Solar Probe (PSP) observations in the innermost section of the HCS ($< 0.1$ AU) provided evidence of particle acceleration in its vicinity: studies by \citet{Desai2022} and \citet{Desai2023} reported energetic ion enhancement, including protons, helium, oxygen, and iron, with energy $\sim 10-100$ keV nucleon$^{-1}$ during PSP's crossings of the HCS. The energy spectra of the ions exhibit power-law trends with spectral indices ranging from $\sim 4$ to $6$.
Additionally, the maximum energies for different ion populations scale as a function of the charge-to-mass ratio, following $E_\text{max} \propto (Q / M)^{\alpha}$ with $\alpha \sim 0.7$ \citep{Desai2022}. During a new crossing on Dec. 22, 2022 (Encounter 14), protons with a maximum energy of $\sim 500$ keV were observed \citep{Desai2023AGU}.

These PSP observations shed new light on particle acceleration in magnetic reconnection in the solar wind.  While in-situ spacecraft observations have reported the presence of reconnection within the HCS \citep{Gosling2007,Gosling2007b,Phan2009,Lavraud2009,Phan2021}, energetic particles associated with magnetic reconnection are rarely seen. Particularly, the landmark study by \citet{Gosling2005} showed the lack of energetic particles near reconnection exhausts in the solar wind near 1 AU. Although there has been some evidence suggesting particle acceleration associated with the HCS \citep{Khabarova2011,Zharkova2015,Zhao2018}, the condition for accelerating particles in the HCS beyond 1 AU remains unclear. It is important to note that the available energy per particle $\sim m_p V_A^2$ ($m_p$ is the proton mass and $V_A$ is the Alfv\'en speed in the reconnection upstream) is a only few eV at 1 AU or beyond, which limits the efficiency of particle acceleration. In contrast, near the Sun, the available energy per particle can be up to $1 $ keV. Therefore, particle acceleration by reconnection closer to the Sun can be much more efficient. 

Recent studies have significantly advanced our understanding of particle acceleration during magnetic reconnection \citep{Drake2006,Guo2014,Li2021}. Many of them have examined particle acceleration mechanisms in different regions within the reconnection layer, such as near the reconnection $X-$points (e.g., \citealp{Hoshino2001,Drake2005,Fu2006,Oka2010,Egedal2012,Egedal2015,Wang2016}), in contracting plasmoids \citep{Drake2006,Oka2010,Li2017}, and during plasmoid coalescence \citep{Oka2010,Liu2011,Guo2015,Nalewajko2015}. A widely used approach in these studies is particle-in-cell (PIC) kinetic simulations, including full PIC and hybrid PIC (fluid electrons and kinetic ions) simulations. While earlier simulations had difficulties generating a well-defined nonthermal spectrum, recent studies have clearly shown power-law spectra with spectral indices $\sim 4$ in 3D non-relativistic kinetic simulations of reconnection at low plasma $\beta$~\citep{Li2019,Zhang2021,Johnson2022,Zhang2024}. All the species, including electrons, protons, and heavier ions, can be accelerated, as demonstrated by recent 3D simulations \citep{Li2019,Zhang2024}.

Despite the success of PIC simulations, studying energetic particle acceleration in large-scale reconnection is still a major challenge due to the separation in physical scales between the reconnection layer and the typical ion scales. The HCS length scale  \citep{Winterhalter1994,Liou2021} is orders of magnitude larger than the ion inertial length \citep{Perrone2020}, making 
traditional PIC simulations impractical in modeling the entire reconnection region. To address this issue, several models have been developed recently. The \textit{kglobal} model utilizes a multifluid approach in which energetic electrons are approximated using a guiding-center equation \citep{Drake2019,Arnold2021}. While this model includes energetic particle feedback, it does not explicitly account for pitch-angle scattering or the effects of turbulence, which are critical for particle transport. The code also includes a small perpendicular diffusion of particles to avoid numerical instability, which could act like a 3D effect in producing a power-law energy spectrum \citep{Johnson2022}. Another approach is to solve the energetic particle transport equations (e.g., \citealp{Parker1965,Zank2014b,Li2018b}), which have been applied to the study of reconnection acceleration only recently, particularly for modeling particle acceleration and transport during solar flares \citep{Li2018b,Li2022,Kong2019,Kong2022}. In this approach, when the energetic particle distribution is nearly isotropic due to pitch-angle scattering, the primary acceleration mechanism is due to flow compression \citep{Parker1965,Blandford1987}. Meanwhile, PIC simulations have shown that compression energization is the dominant particle acceleration mechanism in the low-$\beta$ and low guide-field regimes during reconnection \citep{Li2018,Du2018}. Therefore, we anticipate that particles can be efficiently accelerated due to flow compression in a large-scale compressible reconnection layer in HCS, especially in the low-$\beta$ and low guide-field regimes~\citep{Birn2012,Provornikova2016,Guidoni2016}.


In this study, to interpret the latest PSP observations of energetic ion acceleration during HCS crossings \citep{Desai2022, Desai2023}, we study the multi-ion-species acceleration by solving the Parker transport equation with MHD simulations of magnetic reconnection. The MHD simulations are in the high-Lundquist-number regime, leading to the formation of plasmoids. Our study includes multiple ion species (H$^{+}$, He$_{4}^{2+}$, O$_{16}^{6+}$ and Fe$_{56}^{14+}$ ions) observed by PSP, and we find power-law energy flux spectra and energy cutoffs that resemble in-situ data for all ion populations.


We organize this paper as follows. Section \ref{sec:method} provides a description of the model and a list of the plasma parameters used in this work. Section \ref{sec:results} presents our findings, including a general illustration of the observed particle acceleration process, a discussion on the role of both parallel and perpendicular diffusion of energetic particles, and multi-species ion acceleration. In Section \ref{sec:discussion}, we compare our findings with spacecraft observations, focusing on the latest findings of \cite{Desai2022} and \cite{Desai2023,Desai2023AGU}.

\section{Method} \label{sec:method}

\subsection{MHD Simulation} \label{subsec:MHD_setup}
We run a 2D MHD simulation of a reconnecting current sheet subject to the plasmoid instability with Athena++ \citep{Stone2008, Jiang2014}. The code solves the resistive MHD equations for a fully ionized plasma, here presented in the conservative form:
\begin{equation}
    \frac{\partial \rho}{\partial t} + \nabla \cdot (\rho \mathbf{v}) = 0,
\end{equation}
\begin{equation}
    \frac{\partial (\rho \mathbf{v})}{\partial t} + \nabla \cdot \Bigg[ \rho \mathbf{vv} + \Bigg( P + \frac{\mathbf{B} \cdot \mathbf{B}}{2} \Bigg) \mathbf{I} - \mathbf{BB} \Bigg] = 0,
\end{equation}
\begin{equation}
    \frac{\partial e}{\partial t} + \nabla \cdot \Bigg[ \Bigg( e + P + \frac{\mathbf{B} \cdot \mathbf{B}}{2} \Bigg) \mathbf{v} - \mathbf{B} (\mathbf{B} \cdot \mathbf{v}) \Bigg] = \nabla \cdot (\mathbf{B} \times \eta \mathbf{J}),
\end{equation}
\begin{equation}
    \frac{\partial \mathbf{B}}{\partial t} - \nabla \times (\mathbf{v} \times \mathbf{B}) = \eta \nabla^2 \mathbf{B},
\end{equation}
\begin{equation}
    e = \frac{P}{\gamma -1} + \frac{\rho \mathbf{v} \cdot \mathbf{v}}{2} +  \frac{\mathbf{B} \cdot \mathbf{B}}{2}.
\end{equation}
In the equations above, $\mathbf{v}$, $P$, $\rho$ and $e$ are the plasma velocity, gas pressure, mass density and total energy density respectively, $\gamma = 5/3$ is the adiabatic index and $\mathbf{B}$ is the magnetic field. The current density $\mathbf{J}$ is determined via $\mathbf{J} = \nabla \times \mathbf{B}$.

The simulation parameters are consistent with the plasma properties observed by PSP during HCS crossings \citep{Desai2022,Phan2022}. The initial values for reconnection magnetic field and number density are identified from in-situ data at the HCS upstream during the encounters E07 and E08 by \cite{Desai2022} and \cite{Phan2022}, and are listed in Table \ref{table:1}. In the MHD run, the total plasma $\beta \sim 0.5$. The simulation is run in a Cartesian computational domain of size $2 \cdot 10^7 \times 10^7$ km$^2$ with an uniform grid in the $x-y$ plane. We normalize the domain size by a reference length $L_0 = 5 \cdot 10^6$ km, so that $L_x = 4$ and $L_y = 2$. The domain is composed by $8192 \times 4096$ grid cells, whose size is $\Delta x = \Delta y = 4.9 \cdot 10^{-4}$ $L_0$. The plasma velocity is normalized by the upstream Alfvén speed, whose value calculated from $B_0$ and $n_0$ ($V_A = B_0 / \sqrt{\mu_0 n_0 m_p} \sim 112$ km s$^{-1}$) is consistent with E08 observation by PSP \citep{Phan2022}. The normalized time is $\tau_A = L_0 / V_A \sim 4.5 \cdot 10^4$ s $\sim 12.4$ hours. All boundaries are set open, in order to allow both plasma and magnetic flux to travel freely through them \citep{Shen2011,Shen2018,Shen2022}. A resistivity $\eta = 10^{-5}$ is used, corresponding to a Lundquist number $S = 2 \cdot 10^5$. 

The initial conditions for the magnetic field are that of a Harris current sheet with the form:
\begin{equation}
    \mathbf{B} = b_0 \tanh \Bigg( \frac{x - x_0}{d} \Bigg) \hat{y},
\end{equation}
where $b_0 = 1$ is the strength of the reconnecting magnetic field (normalized by $B_0$), $x_0 = 2$ is the $x-$position of the current sheet, and $d = 5 \cdot 10^{-3}$ is the thickness of the current sheet, corresponding to $\sim 2.5 \cdot 10^4$ km. At the HCS, the ratio of the guide field $B_g$ over the reconnection field $B_0$ is generally low. In this work we do not employ a guide field, hence $B_g / B_0 = 0$. Similiarly, previous simulations on HCS reconnection, such as the work of \cite{Zhang2024}, were also performed in the low guide field regime. The current sheet width is resolved by $\sim$ 10 grid points. At $t = 0$, the non-dimensional background density $\rho = 1$, while along the current sheet $\rho = 3$. As the total pressure is balanced initially, the plasma temperature $\propto P/ \rho$ is uniform at $t = 0$.

\begin{deluxetable*}{c c c c c c c c}
\tabletypesize{\small}
\tablewidth{0pt} 
\tablecaption{Key physical parameters of the simulation. Upstream magnetic field strength $B_0$, injection energy per nucleon $E_0$, upstream ion number density $n_0$, reference length $L_0$, turbulence correlation length $L_c$, upstream Alfvén speed $V_A$, ion plasma $\beta$ and turbulence variance $\sigma^2$ employed to model plasma conditions at the HCS. $L_0$ and $V_A$ are used to normalize the domain size and set the time scale of the MHD simulation, while $B_0$ sets the magnetic field strength and the magnetic flux perturbation magnitude. $B_0$, $E_0$, $L_c$ and $\sigma^2$ are used to estimate $\kappa_{\parallel}$ in Equation \ref{eq:kappa_parallel}. \label{table:1}}
\tablehead{
\colhead{$B_0$ (nT)} & \colhead{$E_0$ (keV nucleon$^{-1}$)}& \colhead{$n_0$ (m$^{-3}$)} & \colhead{$L_0$ (km)} & \colhead{$L_c$ (km)} & \colhead{$V_A$ (km s$^{-1}$)} & \colhead{$\beta$}& \colhead{$\sigma^2$}
} 
\startdata 
$200$ & 5 & $1.5 \cdot 10^{9}$ & $5 \cdot 10^6$ & $5 \cdot 10^4$ & 112 & 0.5 & 1 \\
\hline \hline
\enddata
\end{deluxetable*}

In order to trigger the tearing instability and plasmoid formation, a small white noise velocity perturbation $v_\text{pert}$ of magnitude $10^{-2}$ and a magnetic flux perturbation of the form
\begin{equation}
    \Phi_z (x,y) = \Phi_{0} b_0 \cos{ \Bigg[ \frac{\pi (x - x_0)}{L_x} \Bigg]} \cos{\Bigg( \frac{2 \pi y}{L_y} \Bigg)},
\end{equation}
and amplitude $\Phi_{0} = 10^{-3}$ are included initially.

\subsection{Solving the Parker Equation} \label{subsec:Parker_eq}

We numerically solve the Parker (diffusion-advection) transport equation in the reconnection region,
\begin{equation}
    \frac{\partial f}{\partial t} + (\mathbf{v} + \mathbf{v}_d) \cdot \nabla f - \frac{1}{3} \nabla \cdot \mathbf{v} \frac{\partial f}{\partial \ln p} = \nabla \cdot (\mathbf{\kappa} \nabla f) + S,
    \label{eq:Parker_eq}
\end{equation}
where the function $f(x_i,p,t)$ is the particle distribution function and has a dependency on the space $x_i$, momentum $p$ and time $t$. In this work, the momentum distribution is assumed to be nearly isotropic. The term $\mathbf{v}$ is the bulk plasma velocity, $\mathbf{v}_d$ is the particle drift with respect to the bulk fluid, $S$ is a source term, 
and $\mathbf{\kappa}$ is the spatial diffusion coefficient tensor.

In this work, the Parker equation is numerically integrated via the Global Particle Acceleration and Transport (GPAT) code\footnote{\url{https://github.com/xiaocanli/stochastic-parker}}. The GPAT code takes the time-dependent magnetic field and flow velocity computed by the MHD simulation in the input, and solves Equation \ref{eq:Parker_eq} by integrating the stochastic differential equations corresponding to the Fokker–Planck form of the Parker transport equation \citep{Ito2004}. The primary acceleration mechanism is adiabatic compression. Past kinetic simulations have shown that this is a good approximation for a weak guide field~\citep{Li2018}, which is suitable for modeling the HCS. More details on the solution of Parker's equation can be found in \cite{Li2018b}.

In the simulations presented in this study, a low-energy particle population with energy $E_0 = 5$ keV nucleon$^{-1}$ is initiated at $t = 0$ throughout the domain. $E_0$ is selected to be larger than the typical thermal energy per particle $\sim m_p V_A^2$ and corresponds to a particle velocity $\sim 8 V_A$, much larger than the typical flow speed in the reconnection region, so the dynamics of energetic particles can be described by the transport equation. This population can be generated through a range of injection mechanisms, as particles enter the reconnection region \citep{Zhang2021,Zhang2024,French2023}.

In this work we estimate that the energetic particle drift has negligible effects on the reconnection-driven compression energization. In the framework of the Parker transport equation, the drift velocity can be estimated as
$v_d \sim \varepsilon/(qBR)$, where $\varepsilon$ is the particle energy, $q$ is the charge, $B$ is the magnetic field and $R$ is a characteristic length scale. If we impose $\varepsilon = E_0$ (5 keV) and $B = B_0$ (200 nT), and we choose $R$ to be equal to the current sheet thickness $d$ ($2.5 \cdot 10^4$ km), the  drift velocity of protons is $ \sim 9.99 \cdot 10^{-5}$ km s$^{-1}$. The overall drift of particles during the entire simulation is on the scale of several kilometers.

\subsection{Diffusion Processes and Turbulence Models} \label{subsec:turbulence_models}
 
The diffusion coefficient tensor is defined by:
\begin{equation}
    \kappa_{ij} = \kappa_{\perp} \delta_{ij} - \frac{(\kappa_{\perp} - \kappa_{\parallel}) B_i B_j}{B^2},
    \label{eq:diffusion_tensor}
\end{equation}
where $\kappa_{\parallel}$ and $\kappa_{\perp}$ are the diffusion coefficients parallel and perpendicular to the magnetic field, respectively. The energy-dependent value of $\kappa_{\parallel}$ can be estimated from quasi-linear theory \citep{Jokipii1971,Giacalone1999}:
\begin{equation}
    \kappa_{\parallel} (v) = \frac{v^{3}}{4 L_c \Omega_{0}^{2} \sigma^{2} \gamma}\csc \Bigg( \frac{\pi}{\gamma} \Bigg) \Bigg[ 1 + \frac{8}{(2 - \gamma)(4 - \gamma)} \Bigg( \frac{\Omega_0 L_c}{v} \Bigg)^{\gamma} \Bigg],
    \label{eq:kappa_parallel}
\end{equation}
where $v$ is the particle speed, $L_c$ is the turbulence correlation length, $\Omega_0$ is the particle gyrofrequency, $\gamma$ is the turbulence spectral index, and $\sigma^2 = \langle \delta B^2 \rangle / B_0 ^2$ is the variance of turbulence. The model employs an isotropic, magnetostatic turbulence to account for the solar wind turbulence near the HCS. For this approximation to be valid, the particle velocity must be much larger than the wave speed, which is the case of our study (as $v_{\text{particle}} \sim 8 V_A$). Observations show that the correlation length in the near Sun space is $\sim 10^4 - 10^5$ km \citep{Zhao2022}. We set $L_c = 5 \cdot 10^4$ km and assume $\sigma = 1$ in the HCS reconnection region.  $\kappa_{\parallel}$ is normalized by $L_0 V_A$.

Because different ion species have different gyrofrequency and interact with turbulence at different scales \citep{Yu2022}, the value of $\kappa_{\parallel}$ depends on the ion charge-to-mass ratio. Therefore, the following expression can be derived from Equation \ref{eq:kappa_parallel} for heavier ions:
\begin{equation}
    \kappa_{X} \approx \Bigg( \frac{Q_X}{M_X} \Bigg)^{\Gamma} \kappa_p,
    \label{eq:gamma_dependency_coefficient}
\end{equation}
where $Q_X$ and $M_X$ are  the charge number and the mass number for the ion species $X$, respectively, and $\Gamma = \gamma -2$. $\kappa_p$ is the diffusion coefficient for protons. For the classical Kolmogorov turbulence, $\gamma = 5/3$ hence $\Gamma = -1/3$. For 5 keV protons, $\kappa_{\parallel} = 7.99 \cdot 10^{12}$ m$^2$ s$^{-1}$, and the normalized $\kappa_{0} = 1.42 \cdot 10^{-2}$.

Uncertainties still exist around the magnitude of $\kappa_{\perp}$ \citep{Li2018b}. Observations of solar particle events do not strongly constrain this parameter, as the ratio $\kappa_{\perp} / \kappa_{\parallel}$ varies by orders of magnitude and can become very large ($\sim 0.25$ reported by \citealp{Zhang2003}, $\sim 0.13 - 1.47$ observed by \citealp{Dwyer1997}). The analytical work presented by \cite{Matthaeus2003}, however, suggests that $\kappa_{\perp} / \kappa_{\parallel} \sim 0.02-0.05$. Since the scattering time scale is usually longer than the gyroperiod, particles move preferentially along the magnetic field than across it, and $\kappa_{\perp}$ is generally much smaller than $\kappa_{\parallel}$. Test-particle simulations show typical values of $0.02-0.04$ that are nearly independent of particle energy \citep{Giacalone1999}. These estimates are in agreement with the smaller ratios found in other observational works (e.g. \citealp{Palmer1982}, who found a ratio $0.022 - 0.083$ at 1 AU, and \citealp{Lockwood1992}, who found a ratio $\geq$ 0.01 for 70 MeV cosmic ray particles). We investigate the effects of $\kappa_{\perp}/\kappa_{\parallel}$ from a value consistent with previous computational works ($\kappa_{\perp}/\kappa_{\parallel} = 0.03$) to a value in the range estimated by observations ($\kappa_{\perp}/\kappa_{\parallel} =0.5$) to determine a parameter space where particle energization matches the energy fluxes observed by PSP. This study involves five runs (Case 1 and Cases 5 to 8), and is presented in Section \ref{subsec:survey_kappa_perp}. 

Other ambiguities arise from the estimate of $\kappa_{\parallel}$. Numerical simulations of cosmic ray transport show negligible deviations of $\kappa_{\parallel}$ from its quasi-linear prediction for isotropic turbulence spectra, but the deviation increases to a factor of $\sim 2$ for composite turbulence spectra \citep{Giacalone1999}, which are typical of the heliospheric environment \citep{Podesta2007,Salem2009,Bruno2014}. Furthermore, \cite{Bieber1994} found discrepancies of both ions and electrons mean free path with their prediction from quasi-linear diffusion, as observed by the \textit{Helios} mission. Given that turbulence varies with the distance from the Sun \citep{Adhikari2015}, and changes between individual regions of the heliosphere, the turbulence properties can be quite different for energetic particles produced by separate events. This suggests we still do not accurately know the diffusion coefficient for particle transport at the HCS, and it is therefore interesting to see how modifying $\kappa_{\parallel}$ changes particle energization. In Section \ref{subsec:kappa_parallel} we study six runs (Case 6 and Cases 9 to 13) where we vary $\kappa_{\parallel}$ for a proton distribution.
 
\section{Simulation Results} \label{sec:results}

\begin{deluxetable*}{c c c c c c}
\tabletypesize{\small}
\tablewidth{0pt} 
\tablecaption{Ion species, $\kappa_{\parallel}$  estimated at the injection energy $E_0$, and $\kappa_{\perp}/ \kappa_{\parallel}$ selected for the simulations of particle acceleration performed with the GPAT code. The related properties of the energy flux spectra (spectral index $\delta$ and energy cutoff $E_{max}$) calculated at $t = 7.5 $ $\tau_A$ are also listed. \label{table:2}}
\tablehead{
\colhead{Case ID} & \colhead{Ion species}& \colhead{$\kappa_{\parallel} (E_0)/(L_0 V_A)$} & \colhead{$\kappa_{\perp} / \kappa_{\parallel}$} & \colhead{$\delta$} & \colhead{$E_{\text{max}}$ (keV)}\\
\colhead{} & \colhead{} & \colhead{} & \colhead{} & \colhead{($10-100$ keV)} & \colhead{}
} 
\startdata 
 1 & H$^{+}$ & $1.42 \cdot 10^{-2}$ & 0.05 & $4.40 \pm 0.04$ & $213 \pm 10$ \\
 2 & He$_{4}^{2+}$ & $1.79 \cdot 10^{-2}$ & 0.05 & $4.78 \pm 0.06$ & $194 \pm 9$ \\
 3 & O$_{16}^{6+}$ & $1.97 \cdot 10^{-2}$ & 0.05 & $5.04 \pm 0.05$ & $132 \pm 7$ \\
 4 & Fe$_{56}^{14+}$ & $2.26 \cdot 10^{-2}$ & 0.05 & $5.23 \pm 0.07$ & $120 \pm 6$ \\
 \hline
 5 & H$^{+}$ & $1.42 \cdot 10^{-2}$ & 0.03 & $3.33 \pm 0.08$ & $364 \pm 16$ \\
 6 & H$^{+}$ & $1.42 \cdot 10^{-2}$ & 0.1 & $6.22 \pm 0.13$ & $98 \pm 5$ \\
 7 & H$^{+}$ & $1.42 \cdot 10^{-2}$ & 0.3 & $7.11 \pm 0.19$ & $39 \pm 2$ \\
 8 & H$^{+}$ & $1.42 \cdot 10^{-2}$ & 0.5 & $7.83 \pm 0.18$ & $39 \pm 2$ \\
 \hline
 9 & H$^{+}$ & $7.10 \cdot 10^{-3}$ & 0.1 & $4.29 \pm 0.04$ & $176 \pm 9$ \\
 10 & H$^{+}$ & $3.55 \cdot 10^{-3}$ & 0.1 & $3.64 \pm 0.04$ & $364 \pm 16$ \\
 11 & H$^{+}$ & $1.77 \cdot 10^{-3}$ & 0.1 & $3.17 \pm 0.03$ & $552 \pm 23$ \\
 12 & H$^{+}$ & $8.87 \cdot 10^{-4}$ & 0.1 & $2.77 \pm 0.02$ & $1094 \pm 40$ \\
 13 & H$^{+}$ & $4.44 \cdot 10^{-4}$ & 0.1 & $2.44 \pm 0.03$ & $2787 \pm 90$ \\
 \hline \hline
\enddata
\end{deluxetable*}

\subsection{MHD Simulation} \label{subsec:MHD_results}

\begin{figure}[ht!]
    \centering
    \includegraphics[width=0.8\columnwidth,clip=true,trim=0cm 15cm 0cm 0cm]{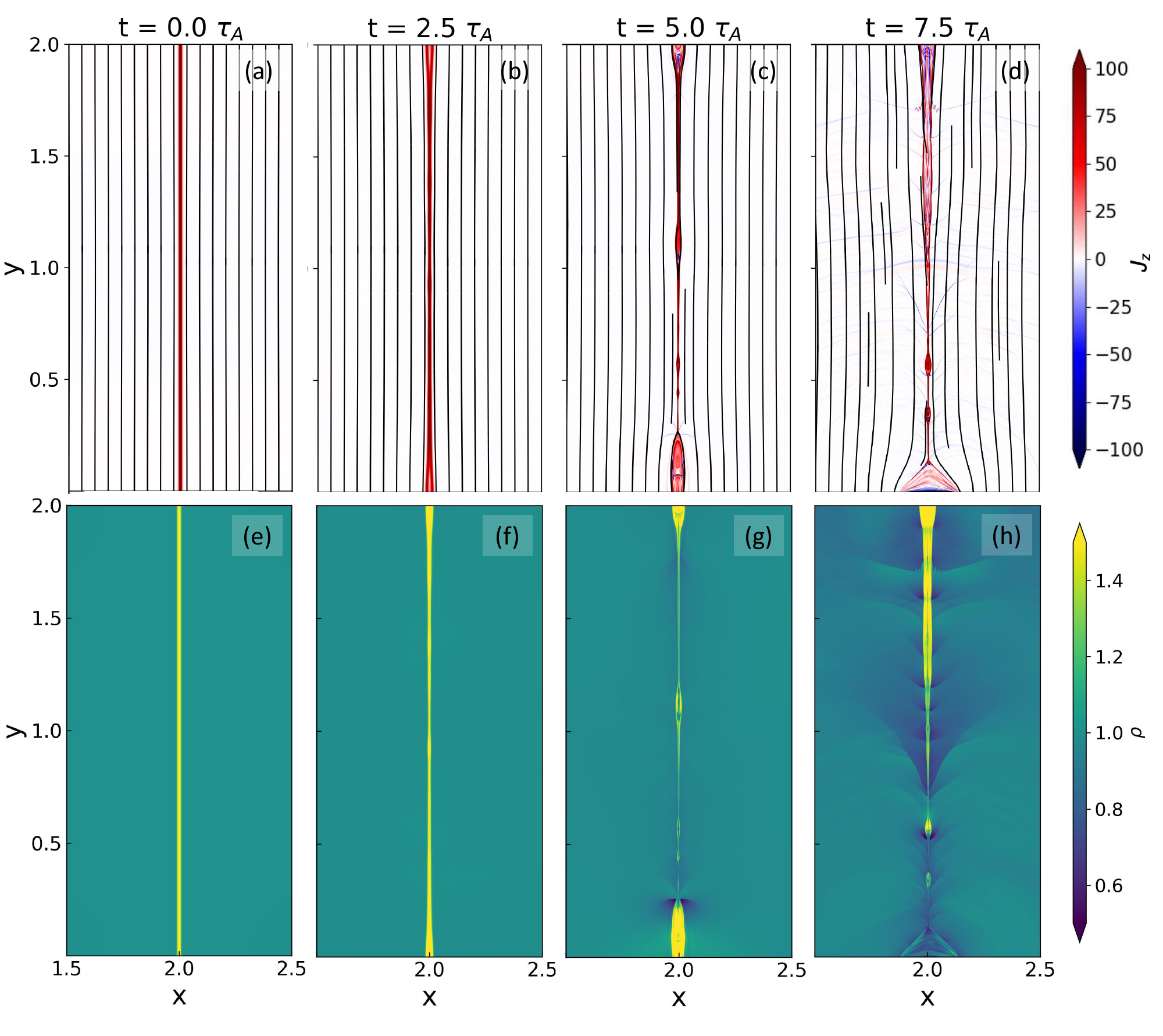}  
    \caption{Out-of-plane current density $J_z$ (top panels) and plasma density $\rho$ (bottom panels) at $t =$ 0, 2.5, 5 and 7.5 $\tau_A$, where $\tau_A$ is the Alfvén crossing time. Magnetic field lines are represented by the black contour lines in panels $(a)$ to $(d)$.}
    \label{fig:1}
\end{figure}

Figure \ref{fig:1} shows the evolution of the out-of-plane current density $J_z$ and the plasma density $\rho$ for the MHD simulation. Starting from the initial setup, the current layer thins due to the perturbation and develops magnetic reconnection with plasmoids. 
Due to the velocity and magnetic perturbations the tearing instability grows, and plasmoids appear around $t = 3$ $\tau_A$. Plasmoids merge before being ejected from the current sheet (see panel $d$). Panels ($c$) and ($d$) show that secondary plasmoids are continuously generated in the unstable current sheet until the end of the simulation.

The density within the current sheet falls from the initial value ($\rho = 3$) to $\rho \sim 1.7$ as the plasma is ejected. The same decrement in plasma density is also found in the plasmoids, with peaks of $\rho \sim 1.8-2$ inside the plasmoids and minima $\rho \sim 0.8$ in the fragments of current sheet immediately outside at $t = 7.5$ $\tau_A$. The density enhancement within plasmoids, compared to the background plasma density, indicates strong compression inside these structures. This is a typical feature of plasmoid-mediated reconnecting current sheets (see, e.g. \citealp{Li2022}).

\subsection{Insights on Particle Acceleration} \label{subsec:acceleration}

\begin{figure}[ht!]
    \centering
    \includegraphics[width=0.98\columnwidth,clip=true,trim=0cm 10.8cm 0cm 0cm]{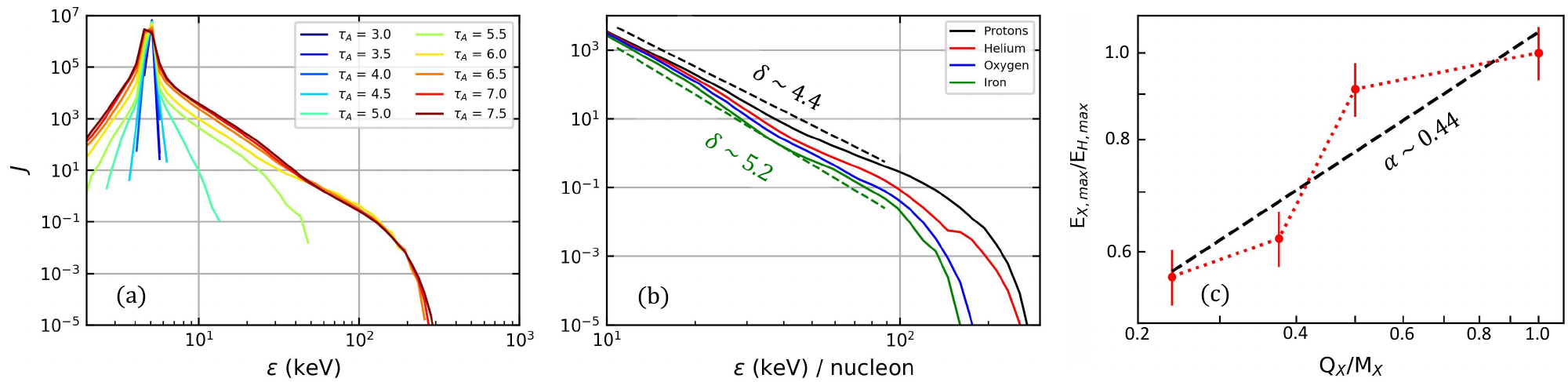}  
    \caption{$a$) Time evolution of the proton energy flux $J$ for Case 1. $J$ is displayed at multiple times for $t = 3.0 -7.5$ $\tau_A$, with a time interval of 0.5 $\tau_A$. $b$) Energy flux (solid lines) for protons (black, Case 1), He (red, Case 2), O (blue, Case 3), and Fe (green, Case 4) in the energy interval $10-300$ keV at $t = 7.5$ $\tau_A$. Dashed lines represent the power-law fit of the spectra for protons (black) and Fe (green), corresponding to the hardest and the softest spectrum of the set, respectively. $c$) Maximum energy per nucleon $E_{X,\text{max}}$ for each species normalized by that of protons ($E_{H,\text{max}}$) as a function of the charge-to-mass ratio $Q_X / M_X$. The black dashed line fits $(Q_X/M_X)^{\alpha}$.}
    \label{fig:2}
\end{figure}

Panel ($a$) of Figure \ref{fig:2} shows how the proton energy flux $J$, calculated over the whole domain as a function of energy, varies with time in the proton reference case (Case 1 in Table \ref{table:2}). Its spectrum follows a power-law distribution, with a well-defined slope at the late stage. We determine the power law spectral index $\delta$ using a linear fit where $\delta  = -d\log J / d\log E$ of the spectrum in the interval $10-100$ keV. The fitting error is calculated as the standard deviation in the same energy range. The proton spectrum, which is a delta function at $t = 0$ $\tau_A$, progressively hardens until $t \sim 6$ $\tau_A$, where the power law trend becomes clear and a well defined energy cutoff appears. After that, $\delta$ becomes nearly constant towards the end of the simulation, varying in the range $\delta \sim 3.9 - 4.4$ in the last ten time steps between $6.5$ $\tau_A$ and 7.5 $\tau_A$. These values are consistent with the measurements of $\delta \sim 4-6$ found in \cite{Desai2022} and \cite{Desai2023,Desai2023AGU}.

Panel ($b$) of Figure \ref{fig:2} shows the energy flux distribution of protons, heliums, oxygens, and irons (Cases 1-4) at $t = 7.5$ $\tau_A$. All four species are accelerated from low energy to well above 100 keV nucleon$^{-1}$ in the nonthermal distribution. We determine the cutoff energy of each species as the energy where the particle spectrum deviates from a power-law by an $e$-fold \citep{Zhang2024}. The spectral indices and cutoff energies of all cases are summarized in Table \ref{table:2}. The proton energy cutoff of Case 1 is at 213 keV; from heliums to irons (Cases 2 to 4), the energy cutoff is found to be in the range $\sim 120-194$ keV/nucleon, decreasing for heavier ions. The PSP observations at the HCS typically show that protons are accelerated to $\sim 100$ keV \citep{Desai2022}, and occasionally to 500 keV (E14, \citealp{Desai2023AGU}). In the same observations, heavier ions tend to have a smaller kinetic energy per nucleon than protons, with the decrease in energy cutoff being dependent on the ion mass (see, e.g. the crossing at E10, \citealp{Desai2023, Desai2023AGU}). This feature is also observed in our simulations: the spectra of heavier ions become progressively softer at the increase of the ion mass, with indices varying from $\delta_{He} \sim 4.8$ for He to $\delta_{Fe} \sim 5.2$ for Fe ions. Once again, these are consistent with the values $\delta \sim 4-6$ reported by \cite{Desai2022} and \cite{Desai2023,Desai2023AGU}.

We further study how the energy cutoff depends on the ion charge-to-mass ratio. PSP data shows a scaling $E_\text{max} \propto (Q / M)^{\alpha}$, where $\alpha \sim 0.6-1.5$ for the crossings in E07-E11. Here we assume that $E_{X,\text{max}}/E_{H,\text{max}}=(Q_X/M_X)^{\alpha}$, where $E_{X,\text{max}}$ is the cutoff energy for the ion species $X$ and is normalized by the proton cutoff energy $E_{H,\text{max}}$. This function is fitted in a log-log space by a straight line: the exponent $\alpha$ is then given by the angular coefficient of the linear fit
\begin{equation}
    \alpha = \frac{d \log (E_{X,\text{max}}/E_{H,\text{max}})}{d \log (Q_X/M_X)}.
\end{equation}
Panel ($c$) in Figure \ref{fig:2} shows the estimated $\alpha$ with the relative error. For the survey in Cases 1-4, $\alpha = 0.44 \pm 0.14$, where the fitting error is calculated as the standard deviation. The errors on the energy ratios (errorbars in panel $c$) are calculated by propagating those associated to $E_{X,\text{max}}$, which correspond to half-width of the energy bin. Our estimates suggest a somewhat smaller $\alpha$ than those observed by \cite{Desai2022} and \cite{Desai2023}.

\subsection{The Role of Perpendicular Diffusion} \label{subsec:survey_kappa_perp}

We examine five runs of proton acceleration where the ratio $\kappa_{\perp} / \kappa_{\parallel}$ is varied in the interval 0.03$-$0.5 to better compare with the wide range reported by observations. The cases are listed with ID 1, 5, 6, 7, and 8 in Table \ref{table:2}.
\begin{figure}[ht!]
    \centering
    \includegraphics[width=0.9\columnwidth,clip=true,trim=0cm 3.8cm 0cm 0cm]{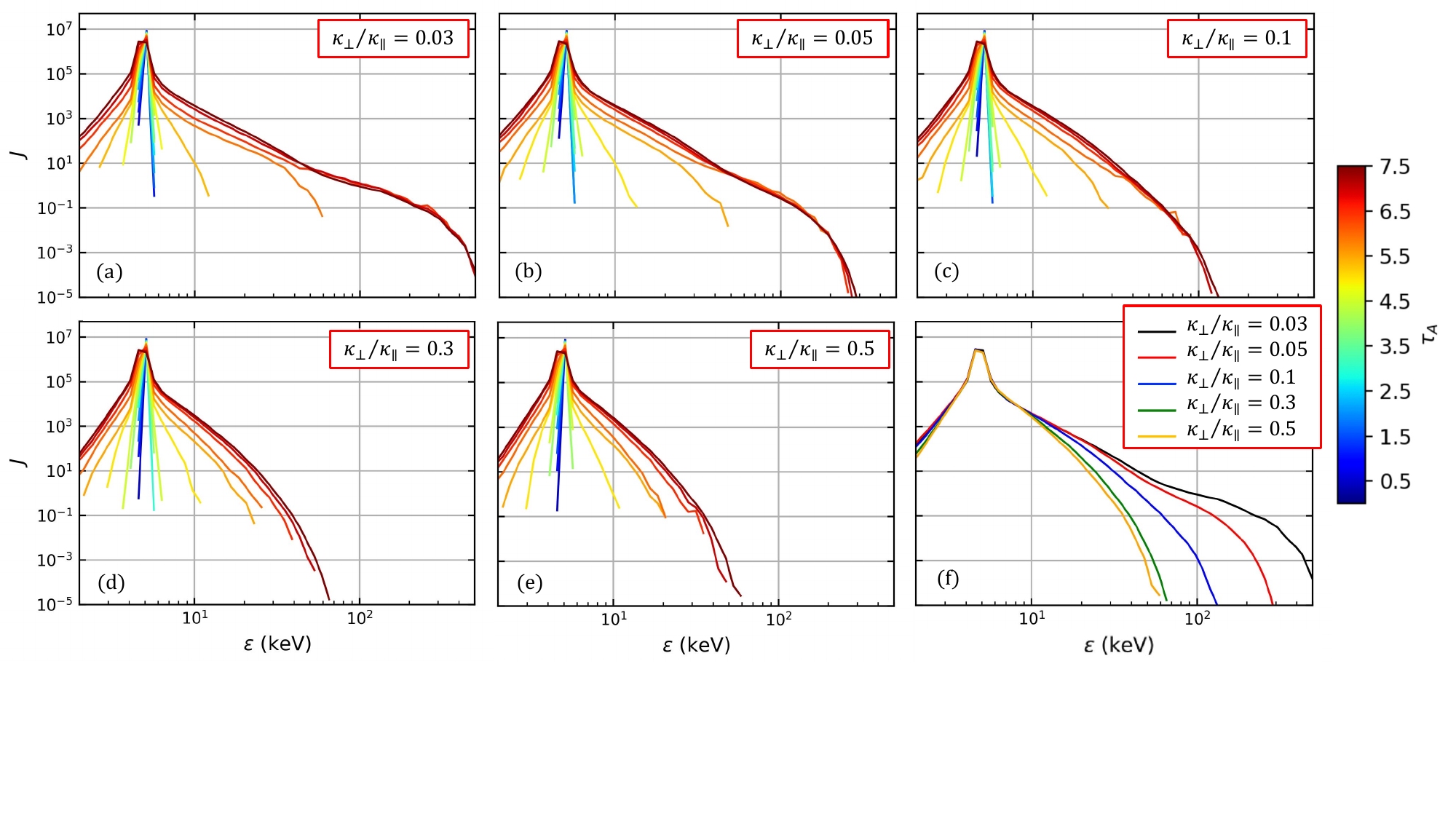}  
    \caption{Proton energy flux varying in time for $a$) Case 5, $b$) Case 1, $c$) Case 6, $d$) Case 7 and $e$) Case 8. The energy spectra are displayed in the interval $0.5 -7.5$ $\tau_A$ with a time step of 0.5 $\tau_A$. The energy spectra of all cases in the survey are plotted at $t=7.5$ $\tau_A$ in $f$).}
    \label{fig:3}
\end{figure}

Figure \ref{fig:3} shows the proton energy flux in the time interval $0 - 7.5$ $\tau_{A}$. As $\kappa_{\perp} / \kappa_{\parallel}$ decreases (panels $a$ to $f$), particles diffuse slower perpendicularly to the magnetic field, and are therefore more confined in the reconnection layer. This results in a stronger acceleration and a harder spectrum, with the power law portion of the spectrum extending over a larger range of energies and a larger cutoff energy.

\begin{figure}[ht!]
    \centering
    \includegraphics[width=0.9\columnwidth,clip=true,trim=0cm 6.5cm 0cm 0cm]{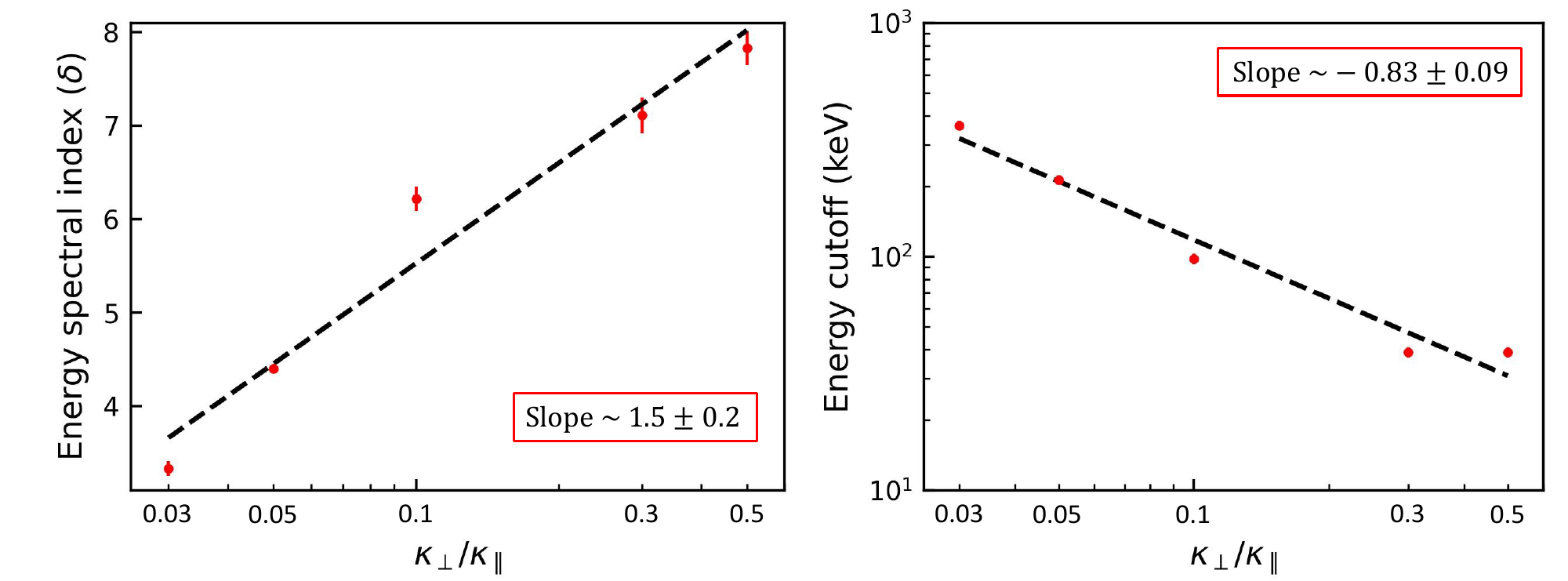}  
    \caption{Spectral index $\delta$ of the energy flux (left panel) and energy cutoff (right panel) plotted as a function of $\kappa_{\perp}/ \kappa_{\parallel}$ at $t = 7.5$ $\tau_A$ for five proton distributions (Cases 1, 5, 6, 7 and 8). The dashed lines show the linear fitting in the log-linear space on the left and the log-log space on the right.}
    \label{fig:4}
\end{figure}

$\delta$ and $E_{\text{max}}$ are plotted in Figure \ref{fig:4} as a function of the $\kappa_{\perp}/\kappa_{\parallel}$ ratio at $t = 7.5$ $\tau_{A}$. The error on the angular coefficient of both linear fits is calculated as the standard deviation.
The left panel shows that $\delta$, varying in the range $\sim 3.33-7.83$ at the increase of $\kappa_{\perp}/\kappa_{\parallel}$, exhibits a logarithmic trend $\sim (1.5 \pm 0.2)\log_{10}(\kappa_{\perp}/\kappa_{\parallel})$. The energy flux spectrum becomes softer as the ratio $\kappa_{\perp}/\kappa_{\parallel}$ increases, as displayed in panel ($f$) of Figure \ref{fig:3}, and at $\kappa_{\perp}/\kappa_{\parallel} > 0.1$ the power law trend of the distributions is less defined (panels $d$ and $e$ of Figure \ref{fig:3}). The energy cutoff falls in the range $E_{\text{max}} \sim 39-364$ keV. In a log-log space (right panel of Figure \ref{fig:4}), $E_{\text{max}}$ appears to have a linear trend with $\kappa_{\perp} / \kappa_{\parallel}$, suggesting that the energy cutoff has a power-law trend with index $0.83 \pm 0.09$.

\subsection{The Role of Parallel Diffusion} \label{subsec:kappa_parallel}

In this Section, we investigate how deviations from the quasi-linear $\kappa_{\parallel}$ can affect the proton energization. The runs examined here are listed in Table \ref{table:2} with ID 6 and 9 to 13. All runs are performed with the same $\kappa_{\perp}/ \kappa_{\parallel} = 0.1$ to maintain a realistic ratio between the coefficients, whereas $\kappa_{\parallel}$ has been systematically reduced by a factor of two at each run, and varies from $1.42 \cdot 10^{-2}$ (Case 6) to $4.44 \cdot 10^{-4}$ (Case 13).

\begin{figure}[ht!]
    \centering
    \includegraphics[width=0.95\columnwidth,clip=true,trim=0cm 3cm 0cm 0cm]{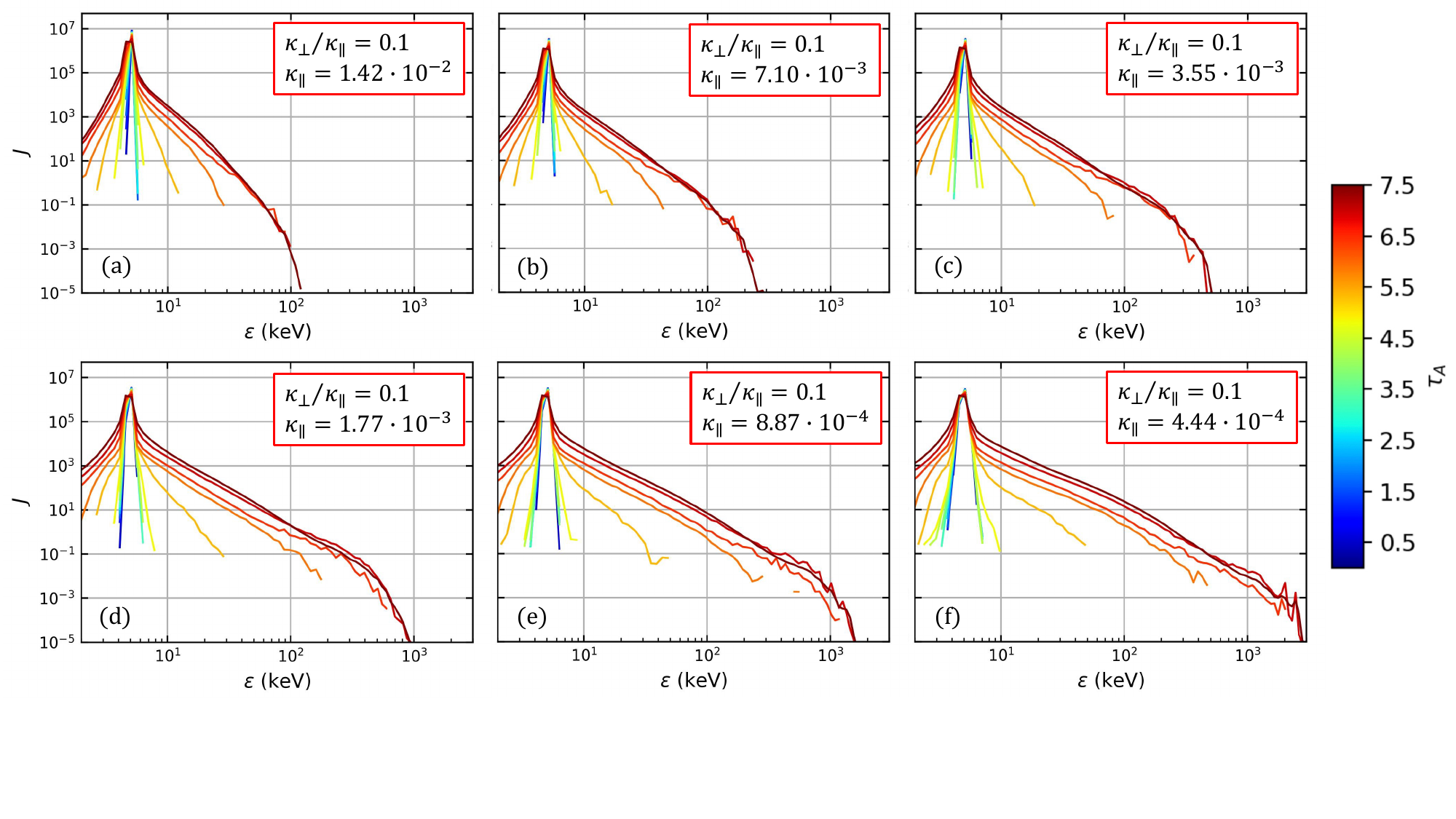}  
    \caption{Proton energy flux varying in time for $a$) Case 6, $b$) Case 9, $c$) Case 10, $d$) Case 11, $e$) Case 12 and $f$) Case 13. The energy spectra are displayed in the interval $0.0 -7.5$ $\tau_A$ with a time step of 0.5 $\tau_A$.}
    \label{fig:5}
\end{figure}
Figure \ref{fig:5} shows the time evolution of the energy flux $J$ for six proton populations. With the decrease of $\kappa_{\parallel}$, and, in proportion, of $\kappa_{\perp}$, particles can be confined in the current sheet longer, yielding more efficient acceleration to higher energies and the formation of harder energy spectra. At the largest $\kappa_{\parallel}$ (panel $a$), the spectrum is soft with $\delta = 6.2 \pm 0.1$, and the power-law trend of the distribution is less clear compared to the other runs, although a flattening can be identified between 20 and 90 keV. Decreasing $\kappa_{\parallel}$, $\delta$ varies in the range $3.17-4.29$ (panels $b$ to $d$) and both a clear power law trend and energy cutoff can be identified. At smaller $\kappa_{\parallel}$ (panels $e$ and $f$), the energy flux spectra show a double power law trend. Case 12 is fitted by two power laws with indices $\delta$ ($10-100$ keV) = $2.77 \pm 0.02$ and $\delta$ ($160-750$ keV) = $3.38 \pm 0.09$. Case 13 can be fitted by two power laws with indices $\delta$ ($10-100$ keV) = $2.44 \pm 0.03$ and $\delta$ ($160-2000$ keV) = $3.75 \pm 0.04$. For these two cases, $E_{\text{max}}$ is obtained by using the energy flux deviation from the second power law calculated in the higher energy range.

\begin{figure}[ht!]
    \centering
    \includegraphics[width=0.9\columnwidth,clip=true,trim=0cm 6.7cm 0cm 0cm]{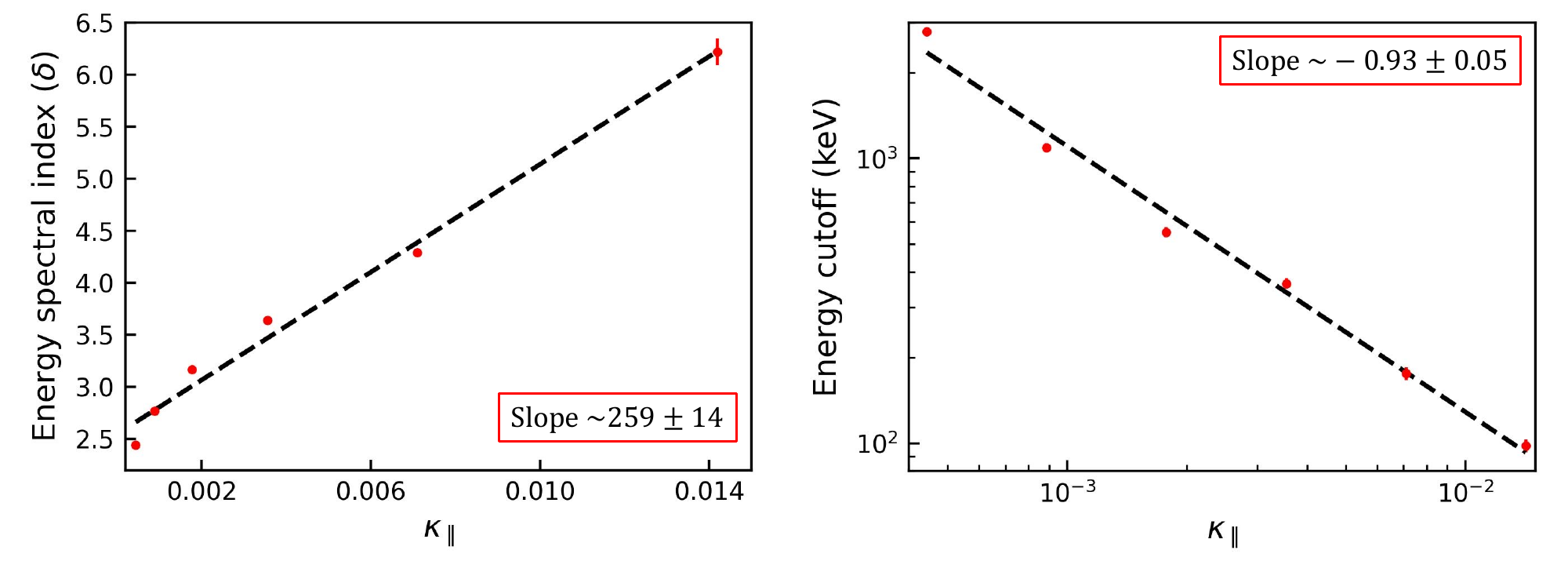}  
    \caption{Spectral index $\delta$ (left panel) and energy cutoff $E_{\text{max}}$ (right panel) plotted as a function of $\kappa_{\parallel}$ at $t = 7.5$ $\tau_A$ for six proton distributions (Cases 6, 9, 10, 11, 12 and 13). The dashed lines show the linear fitting in the linear space on the left and the log-log space on the right.}
    \label{fig:6}
\end{figure}
Figure \ref{fig:6} displays both spectral slope $\delta$ (left panel) and energy cutoff $E_{\text{max}}$ (right panel) varying with $\kappa_{\parallel}$. Both $\delta$ and $E_{\text{max}}$ show a clear trend with $\kappa_{\parallel}$. The spectral index varies linearly as a steep function of $\kappa_\parallel$: the angular coefficient is $259 \pm 14$, where the error is calculated as the standard deviation. Overall, $\delta$ varies in the range $2.44-6.22$. Similarly to what observed for the survey on $\kappa_{\perp}/\kappa_{\parallel}$ in Section \ref{subsec:survey_kappa_perp}, the energy cutoff follows a power-law trend with index $0.93 \pm 0.05$, and varies in the range 98 keV$-$2.8 MeV.

\section{Discussion} \label{sec:discussion}

In this work, we study ion acceleration in HCS near the Sun in a simplified tearing unstable current sheet, which is under plasma conditions consistent with those observed near the Sun. We extensively compare our results with recent observations from PSP \citep{Phan2022, Desai2022, Desai2023, Desai2023AGU}. We estimate the ion spatial diffusion coefficients from quasi-linear theory and perform a parameter survey on both parallel and perpendicular diffusion coefficients $\kappa_{\parallel}$ and $\kappa_{\perp}$. Additionally, we examine the energization of multi-ion species and the dependency on the ion charge-to-mass ratio, comparing it with PSP observations \citep{Desai2022, Desai2023, Desai2023AGU, Zhang2024}. Our main findings are summarized below:
\begin{itemize}
    \item Within the range of diffusion coefficients examined, protons and other ion species can be accelerated into power-law spectra with an energy flux spectral index that varies from $\sim2.4$ to  $\sim7.8$. The spectral index is consistent with the measurements by \cite{Desai2022} and \cite{Desai2023,Desai2023AGU}.
    \item The ion energy cutoff is found to range from $\sim 40$ keV to 2.8 MeV, encompassing the $\sim 10-100$ keV range observed by PSP. The survey on the proton $\kappa_{\parallel}$ reveals cutoff energies that could explain the 500 keV event observed during E14 \citep{Desai2023}. Protons are more efficiently accelerated to higher energies as $\kappa_{\perp} / \kappa_{\parallel}$ decreases, leading to a hardening of the energy spectrum. When $\kappa_{\perp} / \kappa_{\parallel}$ is below 0.1, the maximum energy of the proton distribution matches those in-situ measurements.
    \item When examining the $E_{\text{max}}$ scaling with $Q/M$, we find that $\alpha \sim 0.44$, which is slightly lower than what found by in-situ observations (PSP estimates are in the range 0.63$-1.5$, e.g. \citealp{Desai2022,Desai2023, Desai2023AGU}) and hybrid particle-in-cell simulations by \citealp{Zhang2024}. Further studies should consider different turbulence models and injection processes \citep{Zhang2024} to explain the discrepancy.
\end{itemize}

We find that reconnection acceleration can explain the energetic ions spectral index and energy cutoff observed at HCS crossings by PSP. It must be noted, however, that such match is obtained under different combinations of $\kappa_{\parallel}$ and $\kappa_{\perp}$. Both diffusion coefficients are varied across a wide interval, to take into account the uncertainties on the measurements and potential deviations from theoretical models.

The energy fluxes modeled in this study and those observed by \cite{Desai2022} slightly differ in the dependence on charge-to-mass ratio. Our simulations indicate a clear spectral softening as ion mass increases, whereas in-situ data show that proton spectra are softer than He. In a few cases, proton spectra are also softer than those of heavier ions (O and Fe). This work demonstrates our capability to model the acceleration of different ion species and to obtain results consistent with the limited data available during HCS crossings by PSP. Further detailed studies on $\delta$ and $E_{\text{max}}$ will be necessary as more data from HCS crossings becomes available.

\cite{Desai2022} reports weak time-dependent suprathermal ion anisotropies during the HCS crossings at E07. It is unclear whether these anisotropies are due to the acceleration processes at the source or generated as they propagate out of the acceleration region to where PSP is located. Although the approximation of isotropic distribution still holds in the closest regions to the HCS, addressing the anisotropies more accurately may require modifying the particle transport equation. The GPAT code has been recently upgraded with the Focused transport equation \citep{Zank2014}, which can treat the acceleration and transport of particles with a strongly anisotropic distribution\citep{Zhang2009,Zuo2013,Zhang2017,Kong2022}. We defer a study solving the focused transport equation to a future publication.

\section*{Acknowledgements}

We gratefully acknowledge the helpful discussions with Mihir Desai. G.M. and F.G. acknowledge the support from Los Alamos National Laboratory through the LDRD program and its Center for Space and Earth Science (CSES), DOE OFES, and NASA programs through Grant No.
80HQTR21T0087. X.L. acknowledges the support from NASA through Grant 80NSSC21K1313, National Science Foundation Grant No. AST-2107745, Smithsonian Astrophysical Observatory through subcontract No. SV1-21012, and Los Alamos National Laboratory through subcontract No. 622828. The simulations used resources provided by the National Energy Research Scientific Computing Center (NERSC).

\vspace{5mm}

\software{Athena++ \citep{Stone2008,Jiang2014},  
          GPAT \citep{Li2018b,Li2022}
          }
          
\bibliography{biblio}{}
\bibliographystyle{aasjournal}

\end{document}